\documentclass[reprint,superscriptaddress,showpacs,amsmath,amssymb,aps,prl,floatfix,]{revtex4-1}
\usepackage{graphicx}
\usepackage{dcolumn}
\usepackage{color}
\usepackage{ulem}
\usepackage{bm}
\newcommand{\MG}{Mn$_3$Ge}
\graphicspath{{./}}
\begin{document}
\preprint{APS/123-QED}
\title{Ground state magnetic structure of \MG}
\author{J.-R. Soh}
\affiliation{Department of Physics, University of Oxford, Clarendon Laboratory, Parks Road, Oxford OX1 3PU, UK}%
\author{F. de Juan}
\affiliation{Donostia International Physics Center, 20018 Donostia-San Sebastian, Spain}
\affiliation{IKERBASQUE, Basque Foundation for Science, Maria Diaz de Haro 3, 48013 Bilbao, Spain}
\author{N. Qureshi}
\affiliation{Institut Laue-Langevin, 6 rue Jules Horowitz, 38042 Grenoble Cedex 9, France}
\author{H. Jacobsen}%
\affiliation{Department of Physics, University of Oxford, Clarendon Laboratory, Parks Road, Oxford OX1 3PU, UK}
\author{H.-Y. Wang}
\affiliation{School of Physical Science and Technology, ShanghaiTech University, Shanghai 201210, China}
\affiliation{University of Chinese Academy of Sciences, Beijing 100049, China}
\author{Y.-F. Guo}
\affiliation{School of Physical Science and Technology, ShanghaiTech University, Shanghai 201210, China}
\author{A. T. Boothroyd}
\email{andrew.boothroyd@physics.ox.ac.uk}
\affiliation{Department of Physics, University of Oxford, Clarendon Laboratory, Parks Road, Oxford OX1 3PU, UK}%
\date{\today}
	
\begin{abstract}
We have used spherical neutron polarimetry  to investigate the magnetic structure of the Mn spins in the hexagonal semimetal Mn$_3$Ge, which exhibits a large intrinsic anomalous Hall effect. Our analysis of the polarimetric data finds a strong preference for a spin structure with $E_{1g}$ symmetry relative to the $D_{6h}$ point group. We show that weak ferromagnetism is an inevitable consequence of the symmetry of the observed magnetic structure, and that sixth order anisotropy is needed to select a unique ground state.
\end{abstract}
	
\maketitle

 Recently, \MG\, was found to display a large anomalous Hall effect (AHE) of $\sim$50\,$\Omega^{-1}$cm$^{-1}$ at room temperature~\cite{Kiyohara2016,Nayak2016}. This finding was interesting because Mn$_3$Ge is an antiferromagnetic (AFM) metal, and a large AHE is usually restricted to ferromagnetic metals~\cite{Nagaosa2010}. Moreover, the spontaneous AHE in \MG\, is strongly anisotropic, and can be switched with a small applied magnetic field \cite{Kiyohara2016,Nayak2016}. From a technological standpoint, the concept of an AFM memory device that can be switched is very attractive as there is no demagnetization field, which limits the size of ferromagnetic materials. The prospect of scaling down the size of magnetic devices has prompted many studies of thin-film \MG~\cite{Balluff2018,Kurt2012,Dung2011,Sugihara2015,OGASAWARA2019,Jeong2016}, and the initial results look promising.

Naturally, it is of interest to understand how such a large AHE can occur in an antiferromagnet, and there has been a spate of theoretical studies~\cite{Nayak2016,Guo2017,Kubler2014,Yang2017b,Kubler2018,Yang2017,Liu2017,Ito2017,Zhang_2018,Nyari2019}. The symmetries of non-collinear antiferromagnets generically do not forbid the AHE, and several of the recent studies have concluded that the particular chiral pattern of Mn spins can lead to large Berry curvature at the Fermi surface and thus a large AHE, as predicted by an earlier work~\cite{Hua2014}. The AHE has also attracted recent interest as a signature of Weyl points, which appear relatively near the Fermi level in this system.
%Some of these have suggested that the effect could come from the non-vanishing Berry curvature in momentum space~\cite{Nayak2016,Guo2017,Kubler2014} due to the chiral Mn spin structure, an idea which is inspired by an earlier work~\cite{Hua2014}. Others point to topological Weyl nodes in \MG\, as the source of the large AHE~\cite{Yang2017b,Kubler2018,Yang2017,Liu2017,Ito2017}. 
The theoretical work has led to predictions of other anomalous transport phenomena in \MG, including the anomalous Nernst~\cite{Guo2017}, spin Nernst~\cite{Guo2017} and spin Hall effects~\cite{Nayak2016,Yang2017b,Zhang_2018}. These theoretical predictions depend on the fine details of the magnetic structure, so it is important to work with an unambiguous solution for the zero-field magnetic order.
%Here, we use spherical neutron polarimetry to  obtain an unambiguous solution for the zero-field magnetic structure of \MG.

\begin{figure}[b!]
	\includegraphics[width=0.49\textwidth]{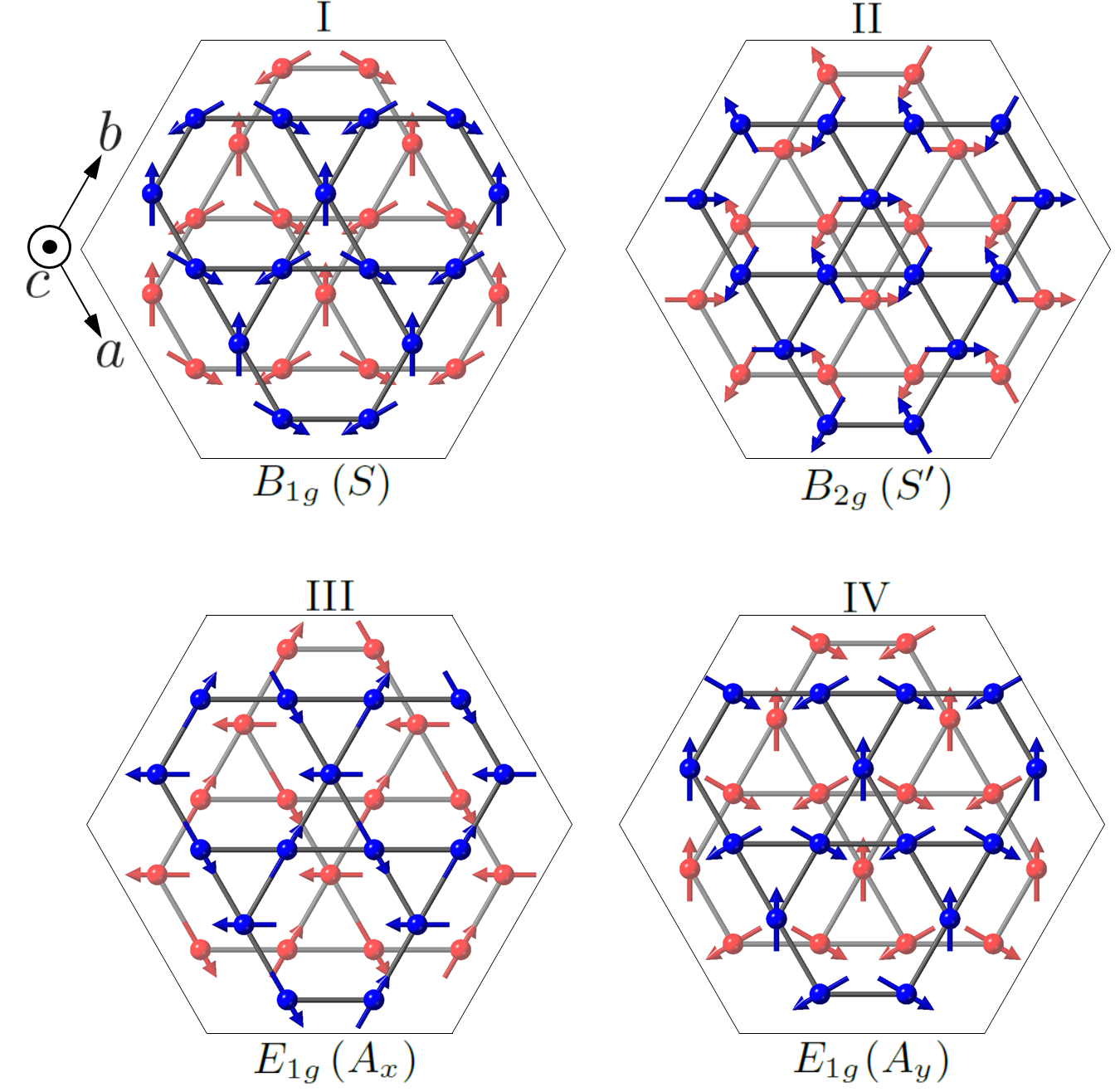}
	\caption{\label{fig:BN4a_Figures_1} Symmetry-allowed magnetic structures of the Mn spins in Mn$_3$Sn/Mn$_3$Ge, viewed in projection down the $c$ axis.  The red and blue arrows correspond to Mn at Wyckoff position $6h$ with $z = 1/4$ and 3/4, respectively.  Ge atoms are omitted for clarity. Only configurations in which the spins related by inversion symmetry are parallel and lie in the basal plane are considered. The structures shown transform according to the irreducible representations (irreps) of the $D_{6h}$ point group. The symmetry label of the irreps is given, together with our labels for the order parameters (in parentheses).  }
\end{figure} 

The hexagonal unit cell of \MG\, can be described by the $P6_3/mmc$ space group (No.~194) with Mn and Ge on the $6h$ and $2c$ Wyckoff sites, respectively. In practice, a small excess of Mn is needed to stabilize the hexagonal phase, so that the true chemical formula is Mn$_{3+x}$Ge$_{1-x}$, with $x = 0.04$ to 0.09 for samples prepared from the melt~\cite{Yamada1988}. For simplicity, we shall continue to write the formula as Mn$_3$Ge. The Mn atoms are arranged in a Kagome pattern, with two Kagome layers per unit cell stacked along the $c$-axis with an in-plane displacement. Antiferromagnetic order of the Mn spins sets in at $T_{\rm N} \simeq 380$\,K, and below roughly the same temperature weak ferromagnetism in the basal plane is observed in magnetisation measurements, with a zero-field remnant moment of about 0.006\,$\mu_{\rm B}$ per Mn at low temperature~\cite{Kouvel1965,Nayak2016,Kiyohara2016,Mn3GeSupp}.

Initial neutron powder diffraction studies of \MG\ in the magnetically-ordered phase revealed that the Mn spins lie in the $ab$ plane in a 120$^\circ$ structure, with a $\textbf{k}=\textbf{0}$ magnetic propagation vector and an ordered moment of about 2.5\,$\mu_{\rm B}$~\cite{Kouvel1965,Kadar1971}. Experiments indicate that the transition to magnetic order in Mn$_3$Ge is second-order~\cite{Kiyohara2016,Nayak2016}, so based on Landau's theory of phase transitions we expect the magnetic structure of Mn$_3$Ge to be described by a single irreducible representation (irrep) of the $D_{6h}$ point group. Symmetry analysis shows that there are four distinct $\textbf{k}=\textbf{0}$ structures with \textit{ab}-plane spin alignment and 120$^\circ$ order (see  Supplemental Material~\cite{Mn3GeSupp}). These are shown in Fig.~\ref{fig:BN4a_Figures_1}. Which of these structures is correct, however, cannot be determined unambiguously from the unpolarized-neutron powder  diffraction data. Subsequently, magnetic diffraction studies were performed on \MG\, single crystals with polarized neutrons~\cite{Tomiyoshi1983,Nagamiya1982}. However, the polarization of the scattered beam, which contains important information for a complete magnetic structure determination~\cite{Brown1991,Brown2001,Qureshi2019,Brown2006} was not analyzed in these experiments. Moreover, the half-polarized  diffraction technique employed in these studies requires the sample to be in an applied field which preferentially orients the Mn moments along the field direction, undermining the elucidation of the true ground state magnetic structure.

To overcome these shortcomings, Brown \textit{et al.}~\cite{Brown1990} used spherical neutron polarimetry (SNP)  --- 
a more sophisticated polarized neutron technique, which probes the sample in zero field (see below) --- to study the magnetic structure of Mn$_3$Sn, which is isostructural to Mn$_3$Ge.  They were able to constrain the spin structure of Mn$_3$Sn to be either model III or IV as shown in Fig.~\ref{fig:BN4a_Figures_1}, but found that both gave an equally good fit to their data~\cite{Brown1990}.

In this work, we used SNP to investigate the zero-field AFM structure of \MG\ by a similar method to that of Brown \textit{et al.}~\cite{Brown1990}. We show unambiguously that the magnetic structure of \MG\, is  described by model IV. 
\begin{figure}[t]
	\includegraphics[width=0.5\textwidth]{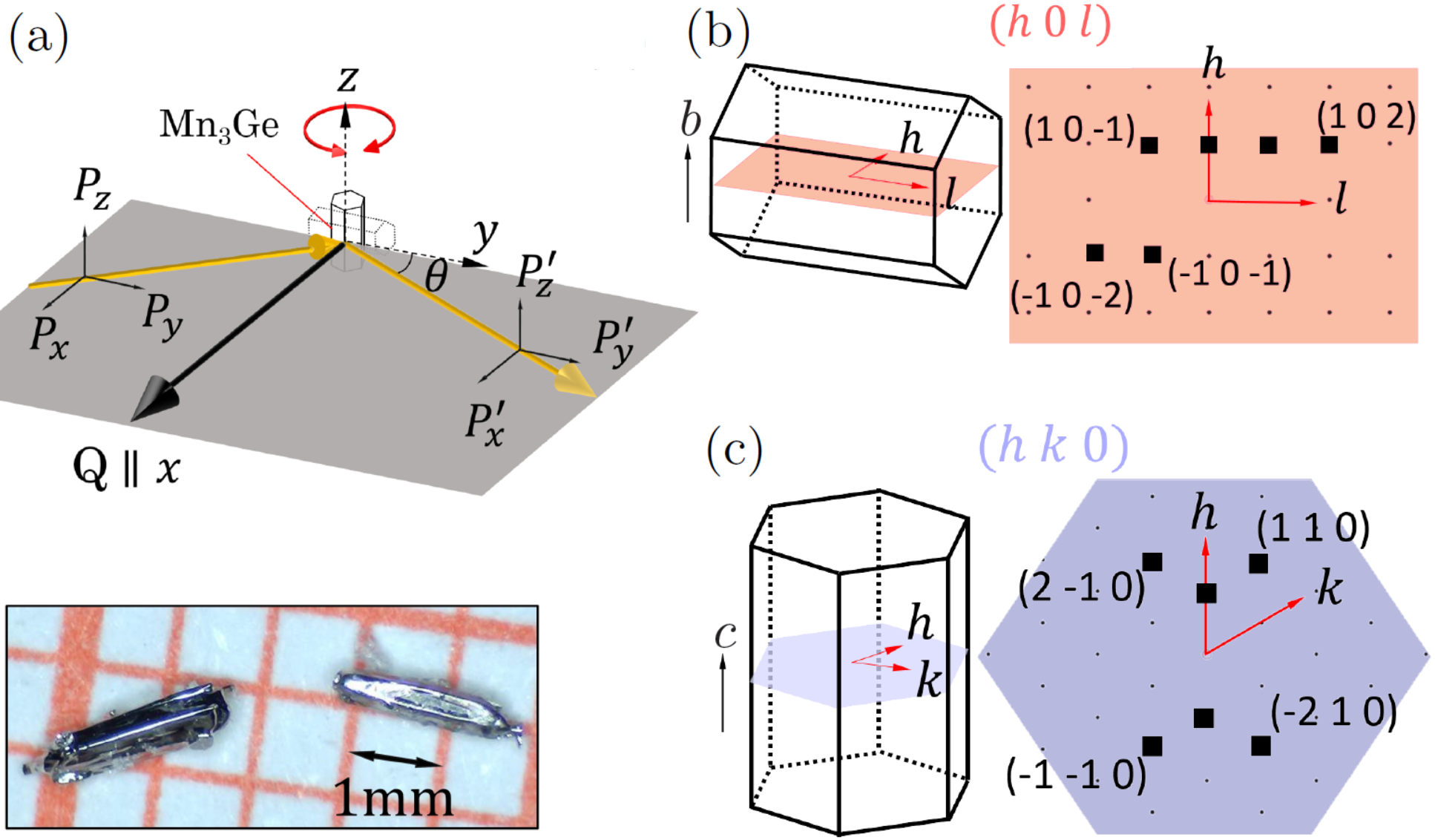}
	\caption{\label{fig:BN4a_Figures_2} (a) The experimental set-up of the SNP of \MG\, in the horizontal diffraction geometry. The photograph shows flux-grown single crystals prepared in this work. (b) and (c) depict the crystal orientations with the $b$- and $c$-axis vertical, respectively, to access the $h0l$ and $hk0$ families of reflections. The reflections studied in this work are labeled with black squares.}
\end{figure}
	
\MG\, single crystals were grown by the flux method. Manganese powder (99.9\%), germanium powder (99.99\%) and cadmium pieces were mixed in a molar ratio of Mn:Ge:Cd = 7:2:48 and placed an alumina crucible. This was sealed in a quartz tube under vacuum and heated to 950$^\circ$C in 5 hours. The temperature was maintained for 20 hours before being slowly reduced to 650$^\circ$C at a rate of 2$^\circ$C/h. The quartz tube was subsequently removed from the furnace to cool to room temperature before being centrifuged to separate the single crystals from the cadmium flux. The flux growth produced shiny metallic needles (see Fig.~\ref{fig:BN4a_Figures_2}) with hexagonal cross-sections and dimensions of up to $2\,\times\,0.4\,\times\,0.4\,$mm$^3$ (length along the crystal $c$-axis). Single crystal x-ray diffraction patterns obtained from the crystals are consistent with the $P6_3/mmc$ space group and demonstrate that the crystals are of good crystalline quality, and the magnetic behaviour is consistent with previous data on Mn$_3$Ge (see Supplemental Material~\cite{Mn3GeSupp}). 
%Moreover, the lattice parameters $a$\,=\,5.320(3)\,\AA\, and $c$=\,4.262(18)\,\AA\, {\color{red}These differ from the values quoted in the supplemental material, and are systematically smaller than reported in the previous studies} are in good agreement with the previous studies~\cite{Yamada1988,Kiyohara2016,Qian2014}.

SNP measurements were performed with the CRYOPAD device installed on the D3 diffractometer at the Institut Laue--Langevin (Grenoble, France), with the sample contained in a zero-field chamber~\cite{Lelievre2005}. The technique involves determining the magnitude and direction of the polarization of the scattered neutrons when the incident neutrons are polarized along each of the principal directions $x$, $y$ and $z$, where $x$ is along the scattering vector $\textbf{Q}$, $z$ is perpendicular to the scattering plane, and $y$ is chosen to complete the right-handed Cartesian set [see Fig.~\ref{fig:BN4a_Figures_2}(a)]. The polarization of the scattered neutrons is resolved along the principal directions, giving a matrix $\textsf{P}$ whose elements $P_{ij}$  represent the $j$ component of the scattered polarization for an incident beam polarized in the $i$ direction. A polarized, monochromatic incident beam was produced by diffraction from the $(111)$ planes of a ferromagnetic crystal of Heusler alloy (Cu$_2$MnAl). Nutator and precession fields were used to control the direction of the incident polarization and the direction along which the scattered polarization was analysed. The scattered beam polarization was measured with a $^3$He spin filter. A correction was made for the time decay of the efficiency of the filter based on measurements of a nuclear Bragg reflection with almost zero magnetic component. 
%the magnitude of the neutron polarization measured at the weak magnetic reflection $(3\,0\,4)$. {\color{red} was this reflection used because it is almost pure nuclear scattering?} 
	
\begin{figure*}[t!]
	\includegraphics[width=0.7\textwidth]{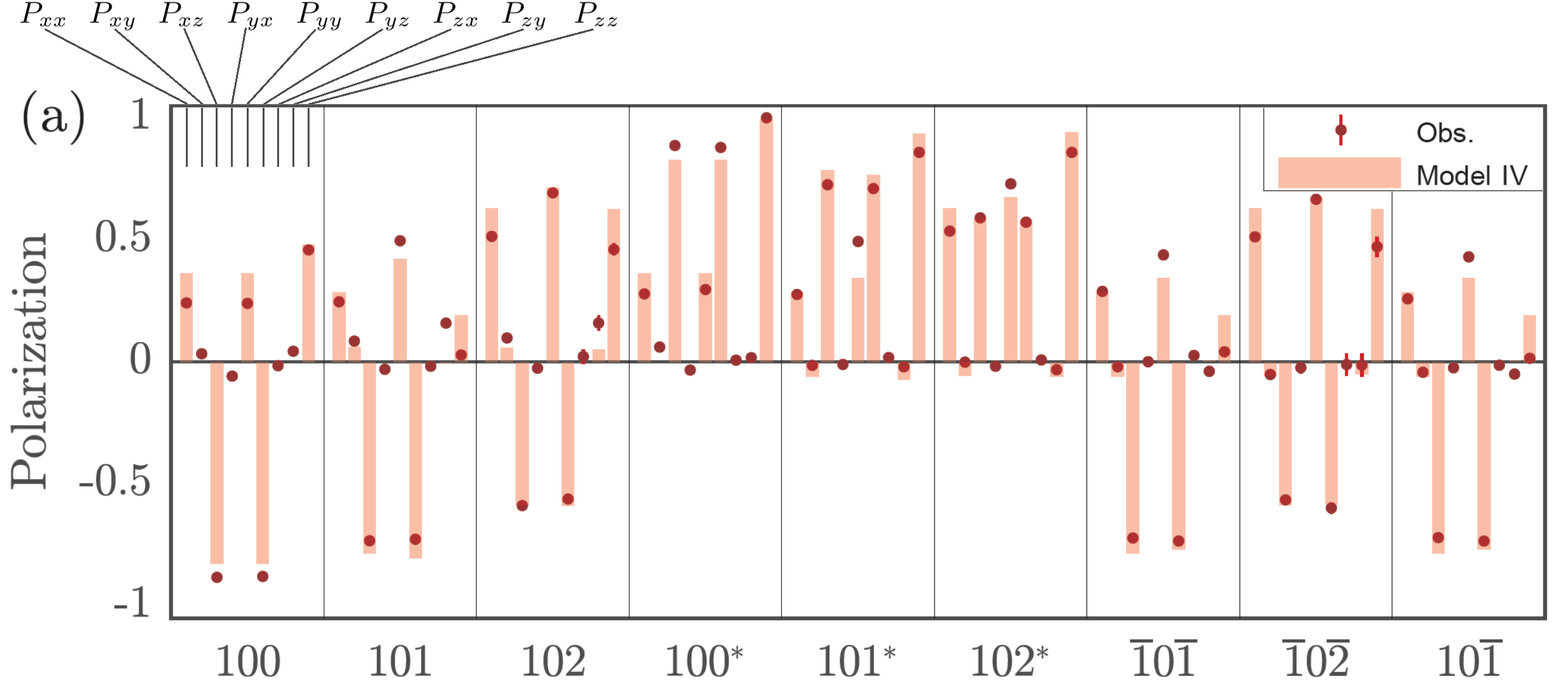}
	\includegraphics[width=0.7\textwidth]{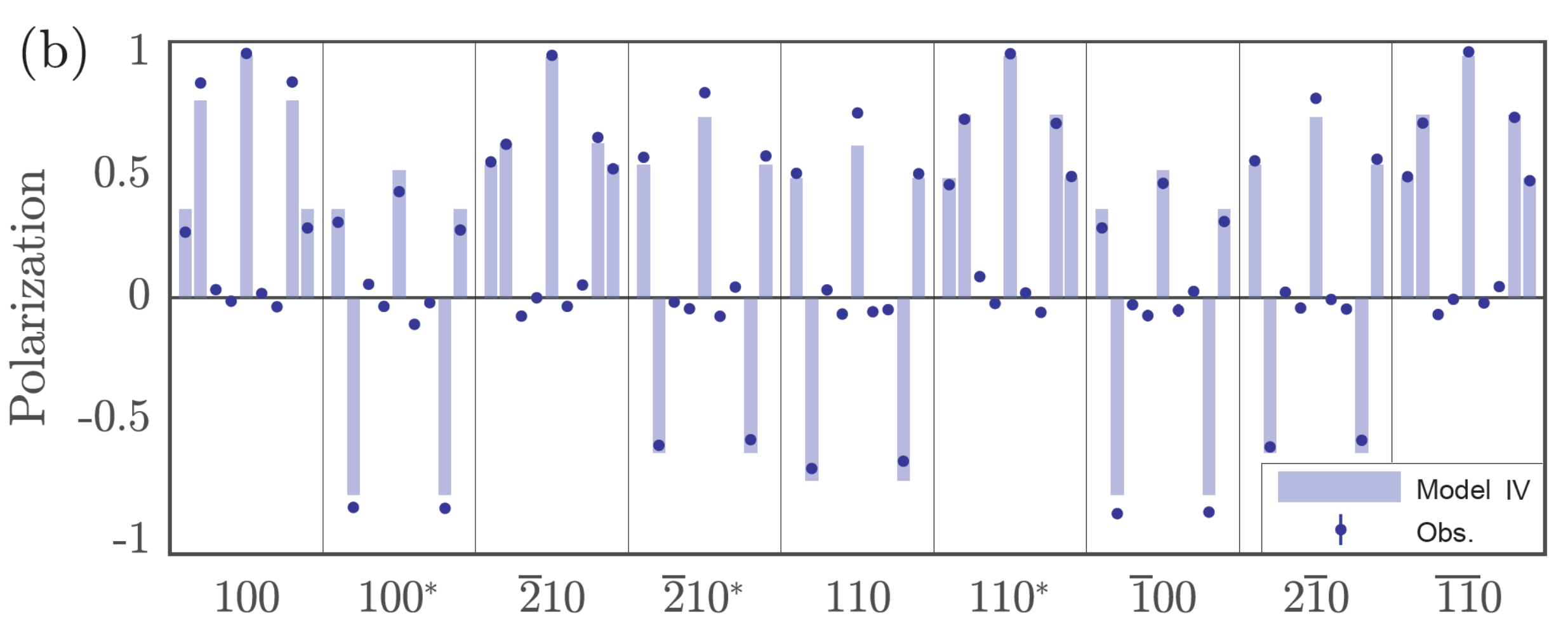}
		\caption{\label{fig:DO5b_Mn3Ge_h0l}	Comparison between the observed and calculated polarization matrix elements $P_{ij}$ for the Bragg peaks measured in the (a) $(h0l)$, and (b) $(hk0)$ scattering planes.  For each reflection, the symbol and vertical bar represent (from left to right) $P_{xx}$, $P_{xy}$, $P_{xz}$, $P_{yx}$, $P_{yy}$, $P_{yz}$, $P_{zx}$, $P_{zy}$ and $P_{zz}$, as indicated. Reflections marked with an asterisk ($\ast$) are measurements that were repeated with the incident polarisation reversed.}
	\end{figure*}

The weak ferromagnetism of Mn$_3$Ge is a potential problem for SNP, as it could cause depolarization of the neutron beam in the sample. We adopted three strategies to alleviate this problem: (1) In the first set of measurements ($h0l$ reflections), the crystal was magnetized in a field of 1\,T applied along the $b$-axis before it was installed in the cryostat mounted on CRYOPAD. This was done in order to reduce depolarization at the boundaries between magnetic domains \cite{Premag}.  (2) The dimensions of the crystal were relatively small, as mentioned earlier. (3) A relatively short neutron wavelength of $\lambda = 0.85$\,\AA\, was used. Depolarization is proportional to the neutron wavelength and the integral of the magnetic flux along the neutron path through the sample. The field integral for a typical path length of 0.5\,mm and remnant magnetization of 0.006\,$\mu_{\rm B}$/Mn is about $3\times 10^{-6}$\,Tm. This corresponds to a maximum neutron precession angle of about $7^\circ$ at $\lambda = 0.85$\,\AA, which can be neglected.

\begin{table}[b]
\caption{\label{tab:table2} The reduced $\chi^2$ goodness-of-fit statistic for the refinements of models I--IV against the measured polarization matrices in the two scattering planes investigated.}
\begin{ruledtabular}
	\begin{tabular}{cccccc}
				&
				\textrm{I}&
				\textrm{II}&
				\textrm{III}&
				\textrm{IV}\\[2pt]
				\colrule
				%$(h0l)$	&261.40 & 1462.15  & 6.88   &  4.14  & \\
				%$(hk0)$	& 118.56 & 204.46    & 118.02  & 1.07  & \\
				$(h0l)$	&21173 & 118434  & 577   & 336  & \\
				$(hk0)$	& 9603 & 165.61   & 9560  & 87  & \\
			\end{tabular}
		\end{ruledtabular}
	\end{table}
	
		The crystal of Mn$_3$Ge was first mounted with the $b$-axis vertical, to access the $h0l$ reflections, and was subsequently remounted with the $c$-axis vertical in order to study the $hk0$ reflections [see Figs.~\ref{fig:BN4a_Figures_2}(b) and (c)]. All measurements were made at a temperature $T=2$\,K.
	
	Figure~\ref{fig:DO5b_Mn3Ge_h0l} presents the set of measured polarization matrix elements $P_{ij}$ for each of the reflections studied [see Figs.~\ref{fig:BN4a_Figures_2}(b) and (c)]. Panels (a) and (b) contain data from the $(h0l)$ and $(hk0)$ scattering planes, respectively. 
	
	For a few reflections, indicated in Fig.~\ref{fig:DO5b_Mn3Ge_h0l} by an asterisk, measurements were made with the incident polarization reversed, as a check. We find that the neutrons suffer from negligible depolarization. This is best exemplified by the matrix elements $P_{zz}$ for the $100^\ast$ reflection in Fig.~\ref{fig:DO5b_Mn3Ge_h0l}(a) and $P_{yy}$ for the $100$, $\overline{2}10$, $110^\ast$ and $\overline1\overline{1}0$ reflections in Fig.~\ref{fig:DO5b_Mn3Ge_h0l}(b), which are all almost unity.
	
	Using the \textsc{Mag2Pol} program~\cite{Qureshi2019}, which is based on the Blume--Maleev equations~\cite{Blume1963,Maleev1963}, we set up the four different magnetic structure models depicted in Fig.~\ref{fig:BN4a_Figures_1}. Where applicable, magnetic domains were also incorporated in the spin configuration models. For instance, for models III and IV, the spin configurations allow for six orientation domains, related by $\pm 60^\circ$ rotation of all of the in-plane Mn spins about the $c$-axis (although only three matter because the scattering cross-section is invariant under $180^\circ$ rotation of all the spins).
	
	For each model I--IV, we calculated the full set of nine matrix elements for each of the measured reflections, and refined the length of the Mn moment and the domain populations (where applicable) via a least-squares fit to the measured polarization matrices (see Supplemental Material~\cite{Mn3GeSupp}). The data from the $(h0l)$ and $(hk0)$ scattering planes were fitted separately. The SNP technique is generally not sensitive to the length of the moment, but when the magnetic propagation vector is $\textbf{k}=\textbf{0}$, as in Mn$_3$Ge, the length of the moment can be obtained from the nuclear--magnetic interference scattering~\cite{Brown2006}.
	
	The values of the reduced $\chi^2$ goodness-of-fit statistic obtained from the different refined models are tabulated in Table~\ref{tab:table2}. The values are large because the number of data points is large and the experimental errors are small. Additionally, the errors include  counting statistics but exclude any sources of systematic error. 
	
	We find that the measured polarization matrices are described best by model IV, which is better than any of the other models by a significant margin. A comparison between the observed and calculated polarization matrices for model IV is given in Fig.~\ref{fig:DO5b_Mn3Ge_h0l}. The agreement is seen to be very good, with deviations of only a few percent for the majority reflections. We also note that model IV is compatible with the observation of weak in-plane ferromagnetism, because only models III and IV allow a weak in-plane ferromagnetic moment while retaining the symmetry of the magnetic structure (see below).
	
	The  estimated  moment  length  is  2.65(2)$\mu_\mathrm{B}$, which is in agreement with earlier studies~\cite{Sukhanov2018,Zhang2013}. Moreover, the domain populations that give the best fit to the data are 60\%, 36(3)\% and 4(1)\%, respectively. The significantly larger population of one domain over the other two in-plane orientations of Mn spins is consistent with the sample having been cooled from room temperature in a 1\,T in-plane field. %The estimated domain populations for the data obtained in the $(hk0)$ scattering plane are 42.3\%, 14(2)\% and 44(5)\%.  The domain populations are relatively more equal as the sample was not cooled in a field for this scattering geometry.
	
There have been a number of attempts to determine the magnetic ground state of Mn$_3$Ge by \textit{ab initio} density functional theory (DFT), with differing results~\cite{Kubler2014,Nayak2016,Zhang2013,Yang2017b,Kubler2018,Yang2017,Guo2017}. 
References~\onlinecite{Nayak2016,Zhang2013,Yang2017b,Kubler2018,Yang2017} predict that the most stable spin configuration is model IV, consistent with our findings. On the other hand,  Ref.~\onlinecite{Guo2017} found the most stable magnetic structure to be model III, and Ref.~\onlinecite{Kubler2014} suggests that the Mn moments display non-planar order~\cite{Kubler2014}. The difference in energy between models III and IV, which are related by an in-plane rotation of  the Mn moments by 90$^\circ$, is reported to be only a few meV, and at the limit of computational uncertainty of DFT ~\cite{Guo2017,Kubler2014,Kubler2018,Nayak2016}.  Moreover, owing to strong electronic correlations among the Mn $3d$ states, the electronic bands near the Fermi level are highly broadened, as also found in Mn$_3$Sn~\cite{Kuroda2017}, making it  difficult to ascertain which  calculation best describes the band structure through comparison with angle-resolved photoemission spectroscopy.  These problems emphasize that, as far as magnetic structure determination is concerned, \textit{ab inito} studies are no substitute for experiment.

In order to understand certain aspects of the magnetic behavior we consider the effective spin Hamiltonian~\cite{Sticht_1989,Liu2017,Nagamiya1982,Nagamiya1979,Tomiyoshi1983,ZImmer1973},
\begin{equation}
\mathcal{H}=\mathcal{H}_\textrm{H}+\mathcal{H}_\textrm{DM}+\mathcal{H}_\textrm{anis},\label{eq:H1}
\end{equation}
where $\mathcal{H}_\textrm{H}$ describes nearest-neighbor Heisenberg exchange,  $\mathcal{H}_\textrm{DM}$ is the in-plane Dzyaloshinskii--Moriya (DM) interaction,  and $\mathcal{H}_\textrm{anis}$ is the orthorhombic single-ion anisotropy. We make the assumptions (based on experiment) that the spins lie in the plane and that spins in one layer in the unit cell are parallel to those in inversion-related sites in the adjacent layer (see Fig.~\ref{fig:BN4a_Figures_1}). The dependence of the Hamiltonian on the active degrees of freedom is then conveniently expressed in terms of four symmetry-adapted order parameters $S$, $S'$, $\bf A$ and $\bf M$, which transform according to irreducible representations (irreps) of the point group $D_{6h}$ (See Fig.~\ref{fig:BN4a_Figures_1} and Supplemental Material~\cite{Mn3GeSupp}). The first two transform as scalars under rotations, and have $B_{1g}$ and $B_{2g}$ symmetry, respectively. ${\bf M} = (M_x,M_y)$, which describes the average in-plane magnetization, and ${\bf A} = (A_x,A_y)$ are 2D irreps  with $E_{1g}$ symmetry. Spin structures III and IV shown in Fig.~\ref{fig:BN4a_Figures_1} correspond to modes $A_x$ and $A_y$, respectively.  

Explicit expressions for the order parameters are given in the Supplemental Material~\cite{Mn3GeSupp}, and the Hamiltonian can be expressed in terms of these as
\begin{align}
\mathcal{H}=&-\frac{J_1}{6}(S^2 + S'^2+{\bf A}^2-2{\bf M}^2)\nonumber \\
&+\frac{D }{2\sqrt{3}}(-S^2-S'^2+{\bf A}^2)\nonumber \\
&+\frac{1}{3}\mbox{\Large\{}K_1S^2+K_2S'^2+\frac{K_1+K_2}{2}({\bf A}^2+{\bf M}^2)\nonumber\\&\hspace{23pt}+(K_1-K_2){\bf A}\cdot{\bf M}\mbox{\Large\}}.\label{eq:H2}
\end{align}
Here, $J_1$ is the nearest-neighbor in-plane exchange interaction, $D$ is the DM interaction, and $K_1$ and $K_2$ are anisotropy constants perpendicular and parallel to the local easy axis, respectively.

The observed spin structure (Model IV) belongs to the $\bf A$ order parameter, so assuming the hierarchy of interactions $|J_1| \gg |D| \gg K_{1,2}$ (Ref.~\onlinecite{Liu2017}) we can conclude that $J_1>0$ and $D<0$. Moreover, once $\bf A$ condenses, a small in-plane magnetization becomes inevitable through the coupling term ${\bf A}\cdot{\bf M}$. The weak ferromagnetism observed in Mn$_3$Ge arises, therefore, because the ground state magnetic structure has the same symmetry as $\bf M$. 

The magnetic ground states described by ${\bf A}$ form a one-parameter  manifold ${\bf A}=A(\cos \theta, \sin \theta)$. The Hamiltonian (\ref{eq:H1})--(\ref{eq:H2}) does not favour any particular $\theta$, and hence does not account for why the system selects $A_y$ ($\theta = \pi/2$) as its ground state. Indeed, earlier studies of the spin Hamiltonian of Mn$_3$Ge reported that the inverse triangular spin structure should have no in-plane anisotropy energy up to fourth order~\cite{Kiyohara2016,Liu2017,Nagamiya1982,Tomiyoshi1983}. Anisotropy  can be introduced if we include a sixth order term in Hamiltonian,
\begin{align}
\mathcal{H}_{6} &= C_1(A_x^3-A_xA_y^2)^2+C_2(A_y^3-A_yA_x^2)^2\nonumber\\
& = \frac{A^6}{2}\mbox{\Large\{}(C_1+C_2)+(C_1-C_2)\cos 6\theta\mbox{\Large\}}.
\label{eq:H3}
\end{align}
This term, which has hexagonal anisotropy, splits the degeneracy of the ground state manifold of $\bf A$ into two states, $A_x$ and $A_y$  (see Supplemental Material~\cite{Mn3GeSupp}). Given that the observed ground state magnetic structure is $A_y$, with $\theta = \pi/2$, we expect $C_1-C_2>0$.

In conclusion, we have determined the magnetic structure of Mn$_3$Ge uniquely, and we have demonstrated that the weak in-plane ferromagnetism observed below $T_{\rm N}$ is intrinsic to Mn$_3$Ge and an inevitable consequence of the symmetry of the magnetic structure. We have also shown that the magnetic ground state is selected by sixth-order anisotropy. The results of this work will be important in future theoretical studies which address the discrepancies between the calculated and measured AHE in Mn$_3$Ge~\cite{Kiyohara2016,Nayak2016,Kubler2014,Kubler2018,Ito2017}. 

Neutron diffraction data from this study are available at Ref.~ \cite{Boothroyd2018}.

\textit{Note added.} During review of our manuscript, we became aware of a conventional polarized neutron diffraction study of \MG\, which found the same magnetic structure as presented here~\cite{Chen2020}.

\begin{acknowledgments}
The authors  wish  to  thank S. Vial (ILL), D. Prabhakaran (Oxford) and M. C. Giordano (EPFL) for technical help, and N. Schr\"{o}ter (PSI) for interesting discussions. This work was supported by the U.K.~Engineering and Physical Sciences Research Council (Grant Nos.~EP/N034872/1 and EP/M020517/1), the Natural Science Foundation of Shanghai (Grant No.~17ZR1443300) and the National Natural Science Foundation of China (Grant No. 11874264).  J.-R. Soh acknowledges support from the Singapore National Science Scholarship, Agency for Science Technology and Research.

\end{acknowledgments}
\nocite{*}
\bibliography{library}

\begin{thebibliography}{42}
\bibitem [{Pre()}]{Premag}%
  \BibitemOpen
  \href@noop {} {}\bibinfo {howpublished} {In retrospect, this procedure was
  probably not effective, because for the second set of measurements ($hk0$
  reflections) we did not pre-magnetize the sample, and in neither case did we
  find any evidence for depolarization.}\BibitemShut {Stop}%
\end{thebibliography}
\end{document}